\documentclass[12pt]{article}
\usepackage{psfig}

\def\gtrsim{\mathrel{\mathpalette\vereq>}}
\makeatletter
\def\vereq#1#2{\lower3pt\vbox{\baselineskip1.5pt \lineskip1.5pt
\ialign{$\m@th#1\hfill##\hfil$\crcr#2\crcr\sim\crcr}}}
\makeatother

\begin{document}

\begin{titlepage}
\begin{center}
\today     \hfill    LBNL-42593 \\
~{} \hfill UCB-PTH-98/61  \\
~{} \hfill hep-ph/9812307\\

\vskip .1in

{\large \bf Establishing a $\nu_{\mu,\tau}$ Component \\
in the Solar Neutrino Flux}%
\footnote{This work was supported in part by the U.S. 
Department of Energy under Contracts DE-AC03-76SF00098, and in part by the 
National Science Foundation under grant PHY-95-14797.  HM was also 
supported by the Alfred P. Sloan Foundation and AdG by CNPq (Brazil).}

\vskip 0.3in

Andr\'e de Gouv\^ea and Hitoshi Murayama

\vskip 0.1in

{\em Department of Physics, University of California\\
Berkeley, CA 94720} \\
and

\vskip 0.05in

{\em Theoretical Physics Group, Lawrence Berkeley National 
Laboratory\\ Berkeley, CA 94720}

\vskip 0.05in

\end{center}

\vskip .1in

\begin{abstract}

  We point out that the recoil electron kinetic energy spectra in the
  $\nu$-$e$ elastic scattering are
  different for incident $\nu_{e}$ or $\nu_{\mu,\tau}$, and hence one
  can in principle establish the existence of the $\nu_{\mu,\tau}$
  component in the solar neutrino flux by fitting the shape of the
  spectrum.  This would be a new model-independent test of the
  solar neutrino oscillation in a single experiment, free from
  astrophysical and nuclear physics uncertainties.  For the $^7$Be
  neutrinos, it is possible to determine the $\nu_{\mu,\tau}$
  component at BOREXINO or KamLAND, if the background is sufficiently
  low. Note that this effect is different from the distortion in the
  incident {\it neutrino}\/ energy spectrum, which has been discussed
  in the literature.

\end{abstract}

\end{titlepage}

\newpage
\setcounter{footnote}{0}
\section{Introduction}

The solar neutrino problem, the fact that the detected neutrino flux
from the Sun is less than the predicted flux, has been known for
decades since the pioneering work of R. Davis in the Homestake mine
\cite{Cl}.  Since then, substantial progress has been made.  The
Kamiokande collaboration confirmed that the neutrinos are indeed
coming from the Sun in a real-time experiment with directional
capability \cite{Kamiokande}.  Both the Kamiokande and Super-Kamiokande
collaborations \cite{Super-K} also reported a depletion of the
predicted flux.  The GALLEX and SAGE experiments, which are sensitive
to the (dominant) $pp$ component of the solar neutrino flux \cite{Ga},
directly related to the solar luminosity, also found a depletion of
the predicted flux.  Without relying on the standard solar model
calculations, one can conclude from the data that the electron
neutrino flux from the ${}^{7}\mbox{Be} + e^{-} \rightarrow
{}^{7}\mbox{Li} + \nu_{e}$ is almost totally depleted (see, {\it
  e.g.}\/, \cite{model_indep}).  Furthermore, the credibility of the
standard solar model calculations has been verified by their agreement
with the helioseismology data at better than one percent level 
\cite{Bahcall98}.
These facts amount to strong evidence of new physics in the neutrino
sector, in particular neutrino oscillations.

Even though the evidence for a ``real'' (solar model independent)
solar neutrino problem is very strong, it is not yet completely
established. First, one needs to rely on (at least) two experiments
to conclude that there is,
model independently, a problem.  It would be far more convincing if
one could see a signal of neutrino oscillations in a single
experiment.  Second, all of the experiments have been of the
disappearance type, where one sees a depletion of the predicted flux.
Given the difficulty of neutrino experiments and of theoretical
calculations of nuclear cross sections, an appearance experiment would
be much more convincing evidence of neutrino oscillations.

The SNO experiment \cite{SNO} will go a long way towards resolving
the issues raised above.  It is designed  
to measure the solar $^8$B neutrino flux
via the charged-current (CC) reaction ($\nu_{e} + d \rightarrow e^{-}
+ p + p$) and the neutral-current (NC) reaction ($\nu_{i} + d
\rightarrow \nu_{i} + n + p$, $i=e,\mu,\tau$).  Assuming both of these
processes can be well understood, a difference between the two
measured fluxes would imply that there are neutrinos in the solar
neutrino flux which are not of the electron type; one may even call
this an appearance experiment of $\nu_{\mu,\tau}$.  There is also an
additional oscillation signature in the possible distortion of the
neutrino energy spectrum.  However, if for some astrophysical and/or
nuclear-physics reason the $^{8}$B neutrino flux is lower than
currently predicted, the SNO experiment may be unable to see an
oscillation signature.\footnote{Another possible problem with SNO is
that the CC/NC ratio does not differ from unity if the oscillation is 
into a sterile neutrino.  We will not consider this possibility in this
letter, because a sterile neutrino is theoretically not very natural
(see, however, \cite{Langacker98}).} 
Another possible concern is that the measurement of the
CC/NC ratio involves the separate calibration of the efficiencies in
the CC and NC processes.

On the other hand, if the current data are correct and the solar
neutrinos indeed oscillate (even with an arbitrary $^8$B flux),
there must be neutrinos other than $\nu_{e}$ in the $^{7}$Be neutrino
flux, and their detection would be an unambiguous signal of neutrino
oscillations.  The $^{7}$Be neutrinos will be studied using $\nu$-$e$
elastic scattering at BOREXINO \cite{Borex}, and possibly also at
KamLAND \cite{Kam}, if the background from natural radioactivity can
be sufficiently suppressed.

In this letter, we study the prospect of establishing the
$\nu_{\mu,\tau}$ component of the solar neutrino flux in a completely
solar model-independent analysis.  We point out that the recoil
electron kinetic energy spectrum is different for $\nu_{e}$ and
$\nu_{\mu,\tau}$.  By fitting the shape of the electron energy spectrum,
one can determine the fraction of $\nu_{\mu,\tau}$ in the solar
neutrino flux, without relying on the predicted neutrino flux from the
standard solar model.  We discuss both the $^{7}$Be neutrinos at
BOREXINO or KamLAND and the $^{8}$B neutrinos at Super-Kamiokande or
SNO. This type of model-independent study seems to be difficult with
the $^{8}$B neutrinos at Super-Kamiokande or SNO, but BOREXINO or
KamLAND should have enough statistics to do the analysis with the
$^{7}$Be neutrinos.

The sensitivity to the $\nu_{\mu,\tau}$ component is a strong function
of the $\nu_{e}$ survival probability.  In the parameter range of the
small-angle MSW solution, one can see the $\nu_{\mu,\tau}$ component
of the $^{7}$Be neutrino flux at more than 95\% confidence level with
two years of BOREXINO running, if the background is sufficiently small.
Under the same conditions, the sensitivity at KamLAND would be even
greater.

\section{Electron Recoil Energy Spectra}
\label{sec:spectrum}

The differential cross-section for elastic $\nu_i$-$e$ scattering
($i=e,\mu,\tau$) is known \cite{`tHooft}:
\begin{equation}
        \frac{{\rm d}\sigma_{i}}{{\rm d}y}
        = \frac{2 G_F^2m_eE_\nu}{\pi}\left[g_L^2+g_R^2(1-y)^2
        - \frac{g_Lg_Rm_e}{E_\nu}y \right],
\end{equation}
where $y=T/E_\nu$, $g_L=\sin^2\theta_{W}\pm1/2$ and
$g_R=\sin^2\theta_{W}$. $T=E_e - m_e$ is the recoil electron kinetic
energy and $E_\nu$ is the incoming neutrino energy in the lab frame.
From the kinematics, $y$ is related to the recoil electron scattering
angle $\theta$ by
\begin{equation}
  y=\frac{m_e}{E_{\nu}}
  \left(\frac{2\cos^2\theta} {(1+m_e/E_{\nu})^2 - \cos^2\theta} \right),
\label{angle}
\end{equation} 
and ranges from $y_{min}=T_{{\rm threshold}}/E_{\nu}$ to 
$y_{max}=(1+m_e/(2E_{\nu}))^{-1}$. The sign in the
definition of $g_L$ depends on the flavor of the incoming neutrino: it 
is plus for $i=e$ and minus for $i=a \equiv \mu, \tau$ ($a$ for
active). 

In the presence of oscillations, the $y$ distribution is 
\begin{equation}
        \frac{{\rm d}\sigma_{P}}{{\rm d}y}=
        P \frac {{\rm d}\sigma_{a}}{{\rm d}y}
        +(1-P) \frac{{\rm d}\sigma_{e}}{{\rm d}y}\, ,
\end{equation}
where $P$ is the oscillation probability for $\nu_e\rightarrow\nu_a$. 
Note that ${\rm d}\sigma_P/{\rm d}y={\rm d}\sigma_{e}(\sigma_{a})/{\rm
d}y$ for $P=0,(1)$. 

\begin{figure}[t]
\centerline{
\psfig{file=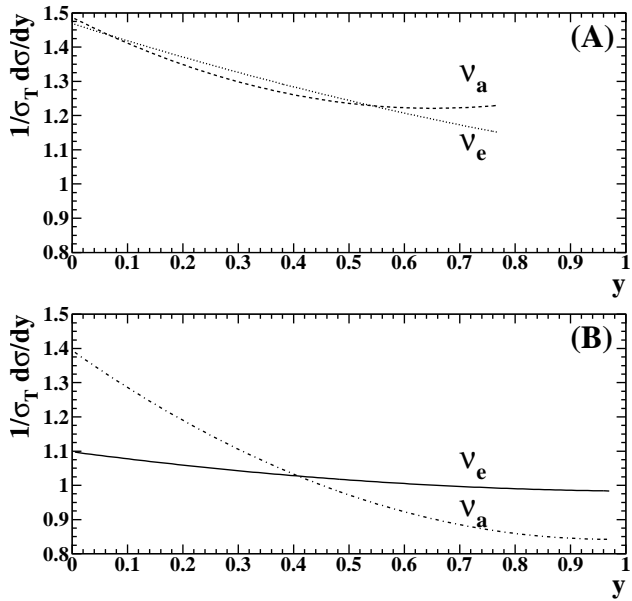,width=0.6\textwidth}
}
\caption{Shape of the recoil electron kinetic energy spectrum from the 
reaction 
$\nu_i+e^{-}\rightarrow \nu_{i}+e^{-}$ as a function of $y=T/E_{\nu}$
for $i=e,a$ and (A) $E_{\nu}=0.862$~MeV or (B) $E_{\nu}=10$~MeV. All
curves are normalized to unit area.}
\label{spectra}
\end{figure}

To illustrate the difference in the recoil electron kinetic energy
spectra for different incoming neutrinos, we plot in
Fig.~\ref{spectra} spectra for two neutrino energies, $E_{\nu} =
10$~MeV (for $^{8}$B neutrinos) and $E_\nu=.862$~MeV (for $^{7}$Be
neutrinos).  The curves are all normalized to unit area such that
their shapes can be compared.  The $\nu_e$ vs $\nu_a$ difference is
more prominent at higher energies, but is not negligible even for the
$^{7}$Be energy.

The central idea of this letter is the following. One should fit the
recoil electron kinetic energy spectrum with an {\it arbitrary
  normalization,}\/ both for $\nu_e$ and $\nu_a$.  The presence of a
non-zero component of $\nu_{a}$-$e$ scattering is the evidence of neutrino
oscillations.  This test does not depend on the theoretical prediction
of the neutrino flux, and hence is independent of solar model and
nuclear physics calculations.  It can be regarded as an ``appearance''
experiment of $\nu_{\mu,\tau}$, similar to the SNO experiment. The
rest of this letter is devoted to discussing under what conditions
such a test can be performed.

\section{$^{7}$Be Neutrinos}

We will analyze the recoil electron kinetic energy distributions for
the case of solar neutrinos produced by the electron capture reaction
$^7$Be $+e^-\rightarrow ^7$Li $+\nu_e$. Because of the $2\rightarrow
2$ kinematics the neutrinos are mono-energetic, which greatly
simplifies our analyses.\footnote{As a matter of fact, there are two
discrete neutrino energies, due to two different final states for
the $^7$Li nucleus, namely $E_{\nu}=0.862, 0.383$~MeV (branching
fractions $90\%$ vs $10\%$).  We focus only on the higher energy
value because the lower energy one does not produce recoil electron
energies above the BOREXINO threshold of 250~keV.}  We mostly focus on
BOREXINO, because it is the only approved experiment which will
specifically study the $^{7}$Be neutrinos.  We do comment on a
possible statistically superior sample from KamLAND.  We find that
BOREXINO can in principle show the existence of a $\nu_{a}$ component
in the $^{7}$Be solar neutrino flux at the two-sigma level after two
years of running, if the background is negligible.

Following the idea presented in the previous section, we will not rely
on the overall rate of the scattering process, which depends on the
theoretical prediction of the flux.  To be completely
model-independent, we use only the shape of the recoil electron
kinetic energy spectrum by allowing the normalization to float in the
fit.  When discussing the sensitivity of BOREXINO or KamLAND, however,
we do need to use some expected neutrino flux; for this purpose, we
use the Standard Solar Model (BP95) \cite{BP95}, 
plus the effect of neutrino oscillations.

The simulated ``data'' sample will consist of ten $y$ bins,\footnote{The
  number of bins is chosen such that the bin size is roughly the same
  as the energy resolution of BOREXINO, so that we do not need to smear
  the energies.} with the number of events in the $k$-th bin given by
\begin{equation}
        N_{k}=
        \frac{N_{T}}{\sigma_{e,T}} 
        \int_{y_{k-1}}^{y_{k}}{\rm d}y\hspace*{1mm}
        \frac{{\rm d}\sigma_{P}}{{\rm d}y},
\end{equation}
where $y_k=y_{min}+(y_{max}-y_{min})k/10$ and
$\sigma_{i,T}=\int_{y_{min}}^{y_{max}}{\rm d}y\hspace*{1mm} \frac{{\rm
    d}\sigma_{i}}{{\rm d}y}$, for $i=e,a,P$.  We take the detection
threshold energy to be 250~keV ({\it i.e.}\/, $y_{min}=0.290$) for
BOREXINO, which is limited by the $^{14}$C radioactivity background.
Note that for the BOREXINO $y$ range, $\sigma_{a,T}/\sigma_{e,T}=0.213$.
$N_{T}=N_{\rm SSM}=55 \times \#\mbox{days}$ is the number of events
predicted by the Standard Solar Model for BOREXINO.  In the
upcoming analysis, we will only consider statistical uncertainties,
and no background.

\begin{figure}[t] 
\centerline{
  \psfig{file=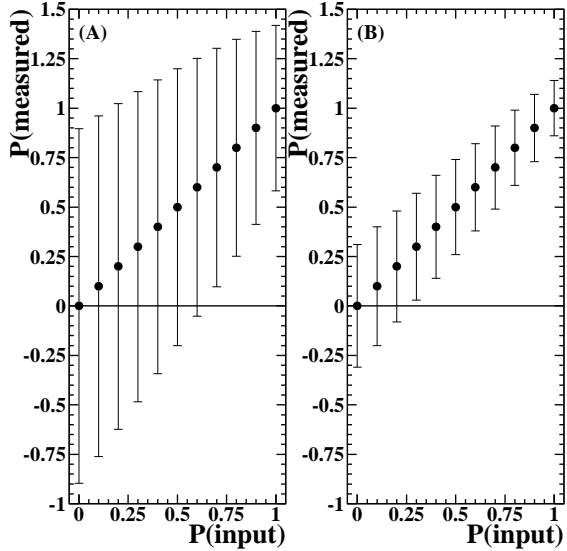,width=0.6\textwidth}
}
        \caption{Measured oscillation probability as a function of the 
          input oscillation probability, in the $\nu_e\rightarrow
          \nu_{a}$ scenario. See text for details. 
          The error bars represent one-sigma statistical 
          uncertainties only. We assume two years of (A) BOREXINO or
          (B) KamLAND running.}
        \label{fit}
\end{figure}

A two-parameter $\chi^2$ fit of the ``data'' events was performed, by
varying both $N_{T}$ and $P$ (two parameters).  This is equivalent to
fitting the data to a linear combination of the $\nu_{e}$-$e$ and
$\nu_{a}$-$e$ differential scattering cross sections with arbitrary
normalizations (two parameters).  Fig.~\ref{fit}(A) shows the extracted
$P_{\rm measured}$ as a function of $P_{\rm input}$ for two years of
BOREXINO running.  A nonzero value of $P_{\rm measured}$ implies the
presence of $\nu_e\rightarrow\nu_a$ oscillations.

The analysis indicates that, for two years of BOREXINO running, the
active neutrino component can be seen at the one-sigma level if
$P\gtrsim 0.7$.  For $P\approx 1$ active neutrino oscillations would
yield more than a two-sigma effect. That is the case for the so-called
small angle MSW solution, which predicts $P\simeq 0.999$.  On the
other hand, the so-called large angle MSW solution predicts $P\simeq
0.50$, and the vacuum oscillations (the ``just-so'' solution) $P\simeq
0.55$ \cite{solar_solution}.

A different type of analysis can be performed, with very similar
results. This different analysis might prove to be useful in order to
deal with the background, if it is not negligible. The integrated
observable $A_1$ is defined by
\begin{equation}
N_{\rm obs}A_1=-\sum_{k=1}^{10}\left(\frac{y_{k-1}+y_{k}}{2}-
\frac{(y_{min}+y_{max})}{2}\right)
  ^{1} N_{k}.
\end{equation}
$N_{{\rm obs}}$ is the number of observed events $N_{\rm
  SSM}\sigma_{P,T}/\sigma_{e,T}$, and the sub(super)script $1$ refers
to the degree of the polynomial multiplying the data. In the absence
of active neutrinos in the solar flux, $A_1=5.79\times 10^{-3}$. Note
that $A_1$ is defined in such a way that the contribution of any
background with a flat $y$ distribution cancels. 

\begin{figure}[t]
\centerline{
  \psfig{file=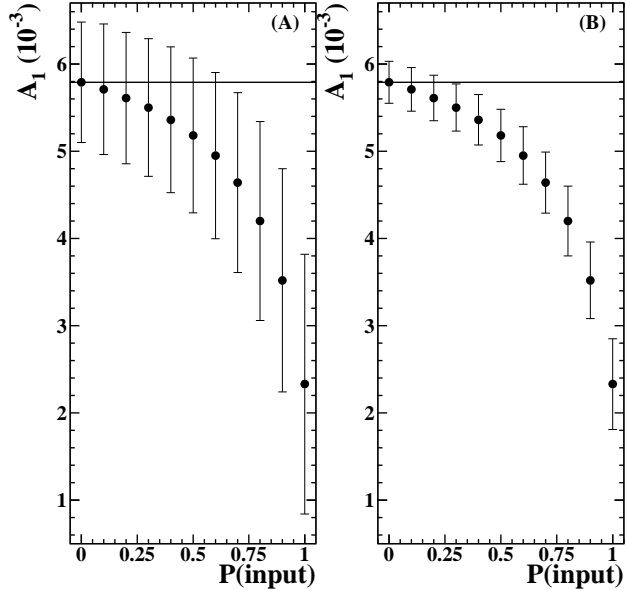,width=0.6\textwidth}
}
        \caption{$A_1$ as a function of the 
          input oscillation probability, in the $\nu_e\rightarrow
          \nu_{a}$ scenario. See text for the definition of
          $A_1$. The horizontal line indicates the value of $A_1$ when 
          there are no active neutrinos (other than $\nu_e$) 
          in the $^7$Be flux. The
          error bars represent one-sigma statistical 
          uncertainties only. We assume two years of (A) BOREXINO or
          (B) KamLAND running.}
        \label{deltaA}
\end{figure}

In Fig.~\ref{deltaA}(A) we plot $A_1$ as a function of $P_{\rm input}$,
for the same conditions considered in the two-parameter fit. The
results are very similar to the ones obtained earlier, as expected.
 
Even though the BOREXINO experiment should have enough statistics for
a model-independent test of the $\nu_{a}$ component in the solar
$^{7}$Be flux, the experimental effort will still be very challenging.
The main concern is radioactive background from Rn, U and Th.
An accurate energy calibration is also crucial.  Our simple analysis is
valid only when the background is sufficiently small in the signal
range.  If the background turns out to be significant, one can still
use the recoil electron kinetic energy spectrum if (1) the background
can be reliably subtracted and (2) the statistical significance can be
kept after the background subtraction.

The first assumption is rather difficult to justify.  The Counting
Test Facility at BOREXINO demonstrated that the background can be
suppressed down to an extremely low level \cite{Borex}, but it was not
possible to prove that it can be suppressed to the required level
because the background was so low that it could not be studied!  Even if
the required level is achieved with the full-scale detector,
understanding the energy spectrum of the background would require a
challenging calibration procedure.

The validity of the second assumption, of course, depends on the level
of the background.  It would be extremely valuable if KamLAND could
also achieve the radio-purity planned for BOREXINO, so that it can
also study the recoil electron energy spectrum from the $^{7}$Be solar
neutrinos, but with a larger fiducial volume.  For comparison, the
same plots as Figs.~\ref{fit}(A) and \ref{deltaA}(A) are shown in
Figs.~\ref{fit}(B) and \ref{deltaA}(B), for two
years of KamLAND running. We assume the BP95 estimate of 466 KamLAND
events per day for a 1~kt fiducial volume.


\section{$^{8}$B Neutrino}

The difference in the recoil electron kinetic energy spectra 
between incident $\nu_{e}$ and $\nu_{a}$ is more prominent for 
$^{8}$B neutrinos than for lower energy neutrinos such as the $^{7}$Be 
neutrinos (see Fig.~\ref{spectra}).
The main complication with the $^{8}$B neutrinos is that, unlike the
$^{7}$Be neutrinos, they have a continuous spectrum.  The
Super-Kamiokande experiment has measured the recoil electron energy
spectrum from $\nu_{i}$-$e$ elastic scattering \cite{SuperK}, which is
a convolution of the neutrino energy spectrum and the $y$ distribution
discussed in Section~\ref{sec:spectrum}.  If the spectrum is not
consistent with expectations, it indicates either that (1) the
neutrino energy spectrum is not the expected one, possibly due to 
unknown nuclear-physics uncertainties in the $^{8}$B beta spectrum (see, 
however, \cite{8B}), (2) the neutrino energy
spectrum is distorted due to an energy dependent neutrino oscillation,
(3) there is some fraction of $\nu_{\mu,\tau}$ in the flux, which
yields a different $y$ distribution, or (4) a combination of them.
The aim of this letter is to identify the possibility (3).

The identification of (3) is, in principle, possible.  If one measures both
the electron recoil energy {\it and}\/ the recoil angle (which is an
observable because we know the direction of the Sun at the time of the
event in a real-time experiment) it is easy to solve the kinematics
and calculate both the incident neutrino energy $E_{\nu}$ and $y$.
Then one can select events with some specific value of $E_{\nu}$ and
study the $y$ distribution.

This program, unfortunately, cannot be done at Super-Kamiokande.  The
main reason is that the recoil angle distribution to too
forward-peaked, $\cos^2\theta\gtrsim 0.9$ from Eq.~\ref{angle}, while
the angular resolution is 25$^\circ$ to 35$^\circ$ in the relevant
energy range \cite{LINAC}.  The strong forward peak happens because of
the high energy threshold for the recoil electron.  Large Time
Projection Chamber (TPC) experiments, such as ICARUS \cite{ICARUS} 
or HELLAZ \cite{HELLAZ}
might have enough angular and recoil energy resolution to attempt such
a program; indeed, HELLAZ quotes a 35mrad ($\sim 2^\circ$) angular
resolution and a $3\%$ $T$ resolution, which is
enough for our purposes.  However, their statistics is very limited
($O(1)$ events per day) and a positive result would require too long a
running time.

SNO studies the recoil electron energy from the charged current
reaction $\nu_{e} + d \rightarrow e^{-} + p + p$, where the energy of
the electron is approximately $T = E_{\nu} + (m_{n} - m_{p} - m_{e}) - B$,
where $B=2.2$~MeV is the deuteron binding energy, when the kinetic 
energy of the recoil protons is neglected.  The measurement of
this recoil electron energy spectrum does not reflect the
$y$ distribution discussed in Section~\ref{sec:spectrum}, but rather
the neutrino energy spectrum.  This is, of course, a very valuable
information in order to study the distortion of the neutrino energy
spectrum due to oscillations.  This is, however, not the effect we
wished to study in this letter.

In principle, one can also try to deconvolute the recoil electron
energy spectrum at Super-Kamiokande using the measured neutrino energy
spectrum from SNO and then determine the presence of a $\nu_{a}$
component in the $^8$B flux via the methods presented in the previous
sections.  As a matter of fact, the SNO experiment itself could also
use the elastic scattering part of its signal to do this analysis.
In principle, SNO could establish active neutrino oscillations even
without its neutron capture capabilities.  This would, however,
require a large elastic scattering sample and hence a very long
running time.

\section{Conclusion}

It seems promising to try to establish neutrino oscillations by
analyzing the recoil electron kinetic energy spectrum in the case of
$^7$Be neutrinos. In particular we have shown that, in the case of
negligible background, two years of BOREXINO running should be enough
to determine the presence of a $\nu_{\mu,\tau}$ component in the solar
neutrino flux model-independently if $P(\nu_e\rightarrow
\nu_a)\sim 1$. Under the same conditions, KamLAND is capable of
obtaining even more significant results.
We emphasize that this effect is unrelated to the distortion of the
incident {\it neutrino}\/ energy spectrum, which has been thoroughly 
discussed in the literature.

It is certainly not clear that the background will be negligible.
Unfortunately we cannot simulate its effects clearly. Instead, we
chose to define two different methods of establishing active neutrino
oscillations. We believe that the background will behave
differently under the two methods, and therefore be more readily
extracted. Another crucial issue is, of course, the energy
calibration. It is clear that a more thorough analysis can only be
performed by detailed simulations of the detectors in questions (and
by the experiments themselves!), which is beyond the scope of our
letter.

Finally, the situation with the $^8$B neutrinos is much less clear,
in part due to their continuous energy spectrum. It is hard to
disentangle distortions in the neutrino energy spectrum, possibly due
to oscillations, from changes in the recoil electron energy
spectrum due to a $\nu_{\mu,\tau}$ component in
the solar flux.  The TPC appears
to be the right technology for this purpose, even though the currently
proposed TPC-based experiments, ICARUS and HELLAZ, 
will not have enough statistics.

\section*{Acknowledgements} 

HM thanks Lawrence Hall, Kevin Lesko, and Stuart Freedman for useful
conversations.  We also thank Stuart Freedman for comments on the
manuscript.  This work was supported in part by the U.S. Department of
Energy under Contracts DE-AC03-76SF00098 and in part by the National
Science Foundation under grant PHY-95-14797.  HM was also supported by
Alfred P. Sloan Foundation, and AdG by CNPq (Brazil).

\end{document}